\documentclass[prl,superscriptaddress,reprint,showpacs,longbibliography]{revtex4-2}
\usepackage{lineno}
\usepackage{color}
\usepackage{physics}
\usepackage{graphicx,textcomp,amssymb,amsmath,dcolumn,hyperref}
\usepackage{siunitx}
\usepackage{comment}

\begin{document}

\title{Three-carrier spin blockade and coupling in bilayer graphene double quantum dots}

\author{Chuyao Tong}
\email{ctong@phys.ethz.ch}
\affiliation{Solid State Physics Laboratory, ETH Zurich, CH-8093 Zurich, Switzerland}

\author{Florian Ginzel}
\affiliation{Department of Physics, University of Konstanz, D-78457 Konstanz, Germany}

\author{Wei Wister Huang}
\affiliation{Solid State Physics Laboratory, ETH Zurich, CH-8093 Zurich, Switzerland}

\author{Annika Kurzmann}
\affiliation{Solid State Physics Laboratory, ETH Zurich, CH-8093 Zurich, Switzerland}
\affiliation{2nd Institute of Physics, RWTH Aachen University, Aachen, 52074, Germany}

\author{Rebekka Garreis}
\affiliation{Solid State Physics Laboratory, ETH Zurich, CH-8093 Zurich, Switzerland}

\author{Kenji Watanabe}
\affiliation{Research Center for Functional Materials, National Institute for Materials Science, 1-1 Namiki, Tsukuba 305-0044, Japan}
\author{Takashi Taniguchi}
\affiliation{International Center for Materials Nanoarchitectonics, National Institute for Materials Science,  1-1 Namiki, Tsukuba 305-0044, Japan}

\author{Guido Burkard}
\affiliation{Department of Physics, University of Konstanz, D-78457 Konstanz, Germany}

\author{Jeroen Danon}
\affiliation{Center for Quantum Spintronics, Department of Physics, Norwegian University of Science and Technology, NO-7491 Trondheim, Norway}

\author{Thomas Ihn}
\author{Klaus Ensslin}
\affiliation{Solid State Physics Laboratory, ETH Zurich, CH-8093 Zurich, Switzerland}

\date{\today}

\begin{abstract}
The spin degree of freedom is crucial for the understanding of any condensed matter system. Knowledge of spin-mixing mechanisms is not only essential for successful control and manipulation of spin-qubits, but also uncovers fundamental properties of investigated devices and material. For electrostatically-defined bilayer graphene quantum dots, in which recent studies report spin-relaxation times $T_1$ up to $\SI{50}{ms}$ with strong magnetic field dependence, we study spin-blockade phenomena at charge configuration $(1,2)\leftrightarrow(0,3)$. We examine the dependence of the spin-blockade leakage current on interdot tunnel coupling and on the magnitude and orientation of externally applied magnetic field. In out-of-plane magnetic field, the observed zero-field current peak could arise from finite-temperature co-tunneling with the leads; though involvement of additional spin- and valley-mixing mechanisms are necessary for explaining the persistent sharp side peaks observed. In in-plane magnetic field, we observe a zero-field current dip, attributed to the competition between the spin Zeeman effect and the Kane--Mele spin--orbit interaction. Details of the line shape of this current dip however, suggest additional underlying mechanisms are at play. 
\end{abstract}

\maketitle
Spin--orbit and hyperfine interactions are common sources of spin decoherence. 
Natural bilayer graphene (BLG) is comprised of $98.9\%$ low-mass, nuclear-spin free $^{\mathrm{12}}$C, and only a small zero-field spin--orbit gap $\Delta_\mathrm{SO}\approx\SI{60}{}-\SI{80}{\mu eV}$ has been experimentally observed~\cite{AnnikaKondo, Luca1001, KaneMele}. Though still in their infancy, investigations of electrostatically defined BLG quantum dots have made great progress, demonstrating high controlability~\cite{MariusPRX, Tunabledot, RebekkaPRL, annikaexcitedstates, AnnikaKondo, Luca1001, Lucacrossover, Samuel2particles, Blockade, SpinT1, SpinT1Aachen}. Recent studies~\cite{SpinT1, SpinT1Aachen} reported spin-relaxation times $T_1$ in BLG quantum dots up to $\SI{50}{ms}$, strongly dependent on magnetic field. The exact spin-mixing mechanisms limiting the lifetimes and determining their magnetic field dependence, however, remain elusive.

\begin{figure}
	\includegraphics[width=8.5cm]{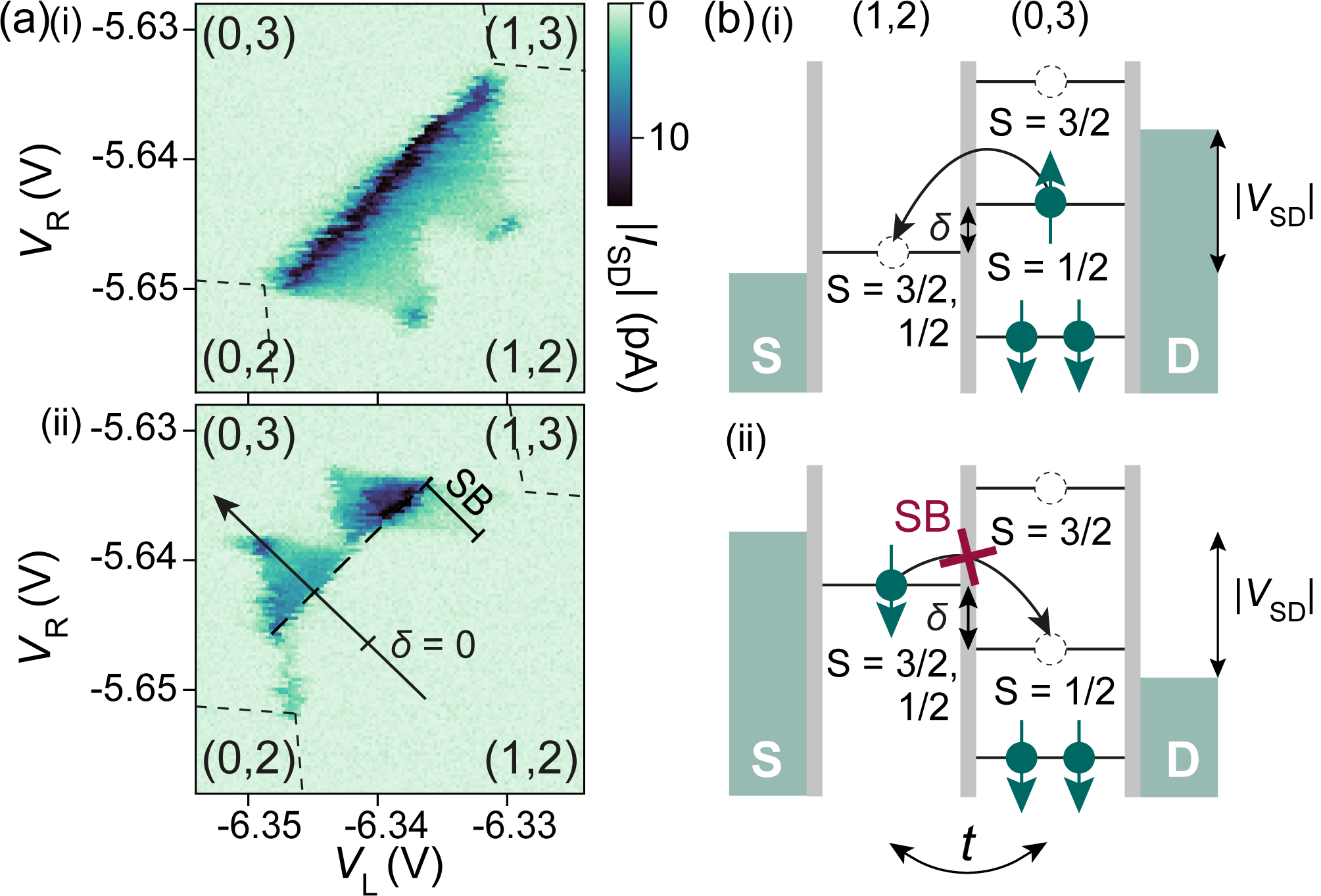}
	\caption{(a) Bias-triangles at zero magnetic field, at $(1,2)$--$(0,3)$ charge degeneracy at $V_\mathrm{MB}=\SI{-6.33}{V}$, (i) $V_\mathrm{SD}=\SI{+1}{mV}$, $(0,3)\rightarrow(1,2)$, and (ii) $V_\mathrm{SD}=\SI{-1}{mV}$, $(1,2)\rightarrow(0,3)$, labeled with the detuning $\delta$-axis being the difference of the $(1,2)$ and $(0,3)$ ground state chemical potential. Regions of strong current suppression induced by spin blockade are marked in (ii). (b) Schematics of $(1,2)\leftrightarrow(0,3)$ charge transport: (i) any $(0,3)$ state can be split into corresponding $(1,2)$ states, but (ii) transport from a spin-polarized $S_z=\pm 3/2$ $(1,2)$ state to $(0,3)$ is spin blocked.}
	\label{figstart}
\end{figure}
A common technique to study spin-mixing mechanisms is examining the dependence of the strength of the Pauli blockade effect on externally applied magnetic field~\cite{koppens2005control, pfund2007suppression, nadj-perge2010, mittag2021few,Li2015,CNTHFNatPhys,jouravlev2006electron,yazyev2008hyperfine, fischer2009hyperfine,danon2009pauli,PhysRevB.86.115322}, where the ``standard" double dot two-carrier charge states $(1,1)$ and $(0,2)$ are investigated [($N_\mathrm{L},N_\mathrm{R}$) labels the number of carriers in the left and in the right dot]. In BLG quantum dots, the valley degree of freedom enriches the energy spectrum~\cite{annikaexcitedstates,Blockade, Samuel2particles, AngelikaQuartetStates, Angelika2021}. Two-electron Pauli blockade of both spin and valley degrees of freedom has been demonstrated, where around zero magnetic field it is mostly only \emph{valley} in nature~\cite{Blockade}, compelling us to move to alternative charge configurations with different ground states to examine \emph{spin}-mixing effects.

We therefore populate our BLG double dots with three electrons, near the $(1,2)\leftrightarrow(0,3)$ charge transition. By close examination of the states involved, we conclude that around zero magnetic field the $(1,2)\to(0,3)$ blockade is truly \emph{spin} in nature. At various interdot tunnel couplings we thus study leakage currents, which exhibit unconventional magnetic field dependencies: In out-of-plane magnetic field, we observe a peak in leakage current around zero field, orders of magnitude too wide~\cite{yazyev2008hyperfine, fischer2009hyperfine,CNTHFNatPhys} to be attributed to hyperfine-induced spin-mixing~\cite{jouravlev2006electron} that is commonly seen in double dots hosted in GaAs~\cite{koppens2005control}, InAs~\cite{pfund2007suppression, nadj-perge2010, mittag2021few}, Si~\cite{Li2015}, and carbon nanotubes~\cite{CNTHFNatPhys}. This wide peak could instead arise from finite-temperature co-tunneling with the leads~\cite{PhysRevB.86.115322,lai2011pauli,PhysRevLett.102.176806}. Sharp side-peaks observed in out-of-plane field point to other spin- and valley-mixing mechanisms at play. In in-plane magnetic field, we observe relatively large leakage current at finite field that dips at zero field. This may arise from the competition between the Kane--Mele spin--orbit interaction polarizing the spins of the blocked states fully out-of-plane~\cite{AnnikaKondo, Luca1001, KaneMele}, and the magnetic field that wants to align the spin quantization axis in plane, thereby mixing blocked and unblocked states~\cite{danon2009pauli}. Compared to similar studies in InAs~\cite{pfund2007suppression,nadj-perge2010}, carbon nanotubes~\cite{CNTHFNatPhys}, and Si~\cite{Li2015,PhysRevB.86.115322} however, not only is the shape of our dip oddly independent of interdot coupling, but it also seems to possess a higher-order magnetic field dependence at small fields, better described by $B^4$, than the typical Lorentzian with $B^2$.

Our double quantum dots are defined electrostatically in the same BLG device as described in Ref.~\onlinecite{Blockade}. Plunger gate voltages for the left and the right dot are $V_\mathrm{L}$ and $V_\mathrm{R}$, respectively. The interdot $t$ and dot-lead tunnel couplings are individually controlled by barrier gate voltages. For details on the sample structure and quantum dot formation, see Appendix~A1. 

Finite-bias transport measurements at zero magnetic field close to the $(1,2)$--$(0,3)$ charge degeneracy are shown in Fig.~\ref{figstart}(a), for (i) ``reverse"-bias, where transport involves the $(0,3)\to(1,2)$ transition, and (ii) ``forward"-bias involving the $(1,2)\to(0,3)$ transition. Strong suppression of current in the lower part of the forward bias triangles signifies the occurrence of Pauli blockade. To understand its nature, we first discuss the relevant single-dot states:

\emph{One particle}.---As shown in Refs.~\onlinecite{AnnikaKondo,Luca1001}, the fourfold degenerate (twofold in spin, $\uparrow$ or $\downarrow$, and twofold in valley, $K^-$ or $K^+$) one-particle ground states in a BLG quantum dot are split by the Kane--Mele~\cite{KaneMele} spin--orbit gap $\Delta_\mathrm{SO}$ into two Kramers pairs: $\ket{\downarrow K^-}$ and $\ket{\uparrow K^+}$ lower in energy, and $\ket{\downarrow K^+}$ and $\ket{\uparrow K^-}$ higher in energy.

\emph{Two particles}.---The two-particle ground state has been consistently observed experimentally to be the threefold degenerate spin-triplet valley-singlet, $\ket{T_s^{\pm(0)}S_v}$ (total spin number $S=1$, where $S_z=-1,0,1$ states are denoted $T^{-,0,+}_s$), due to strong confinements and onsite exchange interaction~\cite{annikaexcitedstates, Blockade, Samuel2particles, AngelikaQuartetStates, Angelika2021}.

\emph{Three particles}.---The three-particle states are most easily understood as being the four different states that can result from removing a single particle with arbitrary spin and valley, from a fully (fourfold) occupied orbital ground state. The resulting spectrum therefore comprises four spin-doublet ($S=1/2$) valley-doublet states, forming Kramers pairs split by $\Delta_\mathrm{SO}$ with $\ket{\downarrow K^-;\uparrow K^+;\downarrow K^+}$ and $\ket{\downarrow K^-;\uparrow K^+;\uparrow K^-}$ lower in energy, and $\ket{\downarrow K^+;\uparrow K^-;\downarrow K^-}$ and $\ket{\downarrow K^+;\uparrow K^-;\uparrow K^+}$ higher in energy. These are the four allowed $(0,3)$ states, with all three carriers on the right dot (shown in Fig.~S2).

We can now investigate the nature of the blockade resulting from the $(1,2)\to(0,3)$ transition. The $(1,2)$ charge states have one carrier in the left dot in any of the four single-particle states, and two carriers in the right dot in any of the three spin-triplet valley-singlet states, forming twelve possible $(1,2)$ states in total. At zero magnetic field these states are split by $\Delta_\mathrm{SO}$, and can be decomposed into four $S=1/2$ spin-doublet states with finite tunneling amplitude to the $(0,3)$ spin-doublets, and eight blocked $S=3/2$ spin-quadruplet states~\cite{PhysRevLett.74.984, PhysRevB.87.241414, PhysRevB.89.085302}, listed in detail in Appendix~A2. The spin--orbit interaction $\Delta_\mathrm{SO}$ mixes the four spin-quadruplet states with $S_z=\pm1/2$ with the spin-doublet states (also with $S_z=\pm1/2$), giving them a finite tunneling amplitude to the $(0,3)$ states.

We are thus left with four truly blocked $(1,2)$ states with $S_z=\pm3/2$, which are simply product states of the relevant one- and two-particle states in the left and the right dot: $\ket{\uparrow K^\pm}_L\ket{T^+_sS_v}_R$ and $\ket{\downarrow K^\pm}_L\ket{T^-_sS_v}_R$. Hence, these states are responsible for the $(1,2)\to(0,3)$ blockade observed in Fig.~\ref{figstart}(a) [as illustrated in Fig.~\ref{figstart}(b)], which is therefore purely \emph{spin} in nature. At large enough detuning, the excited $S_z=\pm3/2$ $(0,3)$ states become accessible and lift the blockade, shown by the finite current reappearing at the tip of the triangles in Fig.~\ref{figstart}(a,ii). 

\begin{figure}
	\includegraphics[width=8.5cm]{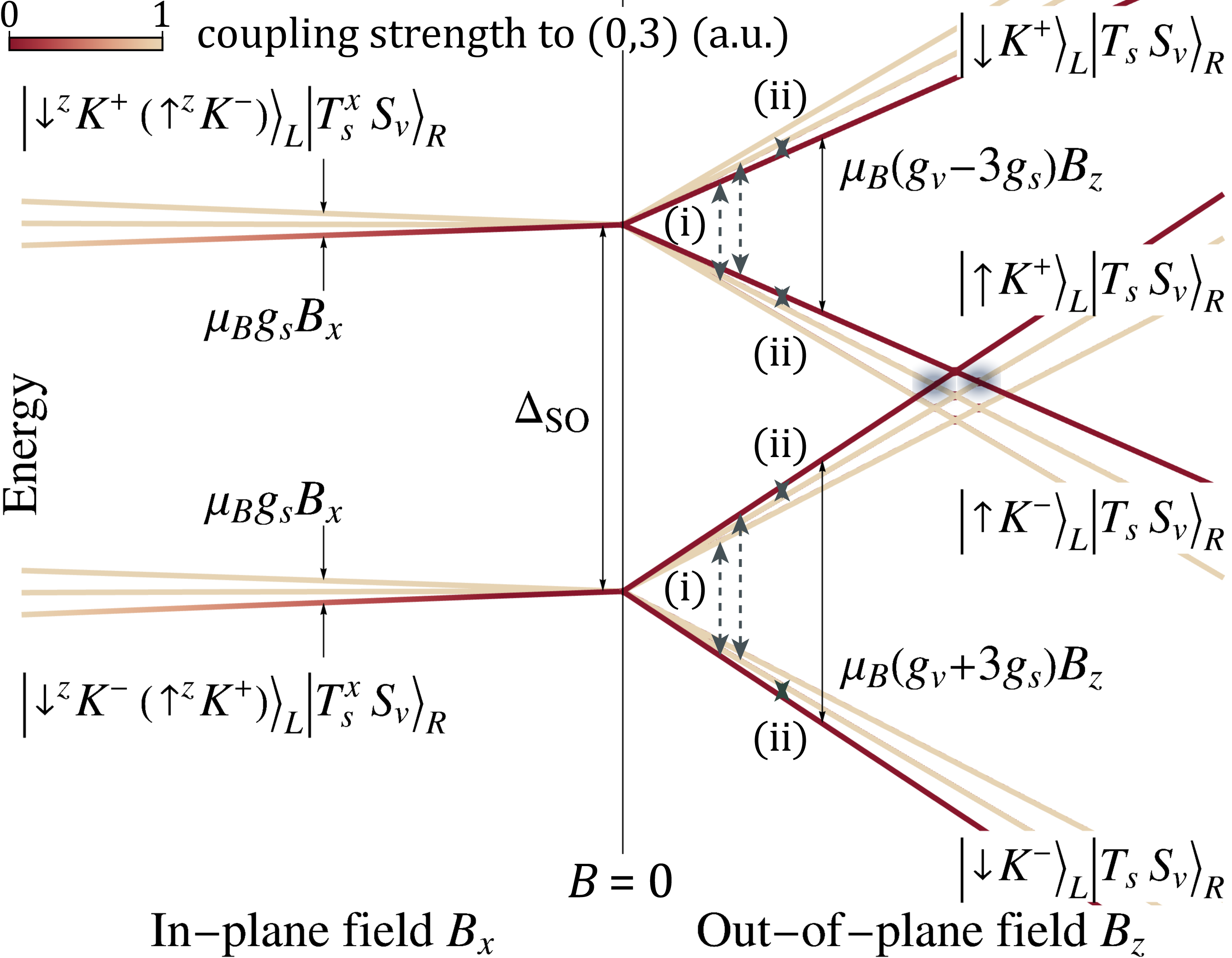}
	\caption{Level structure of $(1,2)$ states as a function of the magnetic field, resulting from the model Hamiltonian (\ref{eq:ham}). The dark red states ($S_z=\pm 3/2$ at $B_x=0$) are fully blocked, while the light states are unblocked. On-site (i) spin- and valley-, and (ii) pure spin-mixing processes can lift the blockade close to $B=0$ by mixing the states pointed at by the arrows, and also near the crossing of different clusters (gray) where $g_v \mu_{\rm B} B_z \approx \Delta_{\rm SO}$. Large enough $B_x$ tilts the spin quantization axis from $z$- to $x$-direction, mixing the blocked states with the unblocked ones, thus lifting the blockade.}
	\label{figlevels}
\end{figure}

To describe the effect of magnetic field on the aforementioned $(1,2)$ states, we employ a simple model Hamiltonian, 
\begin{align}
    \hat H={} & {} \sum_{i}\left(g_s \mu_{\rm B} {\bf B}\cdot \hat {\bf S}_i + g_{v,i} \mu_{\rm B} B_z \hat T_i^z - 2\Delta_{\rm SO} \hat S_i^z \hat T_i^z\right)
    \label{eq:ham}
\end{align}
summed over the left and the right dot $i= L,R$, where $\hat {\bf S}_{i} = \frac{1}{2}\hat{\boldsymbol \sigma}^{(i)}$ is the total spin operator of carriers on the dot $i$ and analogously, $\hat {\bf T}_{i} = \frac{1}{2} \hat{\boldsymbol \tau}^{(i)}$ the total valley pseudospin operator, written in terms of the Pauli matrices $\hat{\boldsymbol \tau}$ that act in valley space. The first term in \eqref{eq:ham} describes the usual Zeeman splitting of the spin states, where the electronic $g$-factor $g_s = 2$~\cite{MariusPRX, RebekkaPRL, annikaexcitedstates, AnnikaKondo, Luca1001}. The second term adds the coupling of the orbital structure of the valley states to the out-of-plane component of the magnetic field; the corresponding valley $g$-factor is displacement field and dot-geometry dependent~\cite{Tunabledot}, measured to be $g_v\approx30$ in this same device at similar gate configuration~\cite{Blockade}. The last term describes the Kane--Mele spin--orbit splitting $\Delta_{\rm SO}\approx 60$--$\SI{80}{\mu eV}$ between states with their $z$-projection of spin and valley aligned parallel or antiparallel~\cite{AnnikaKondo, Luca1001, KaneMele}. These three terms incorporate all the qualitative and partially quantitative understanding of our BLG quantum dot systems to date.

\begin{figure*}
	\includegraphics[width=17.8cm]{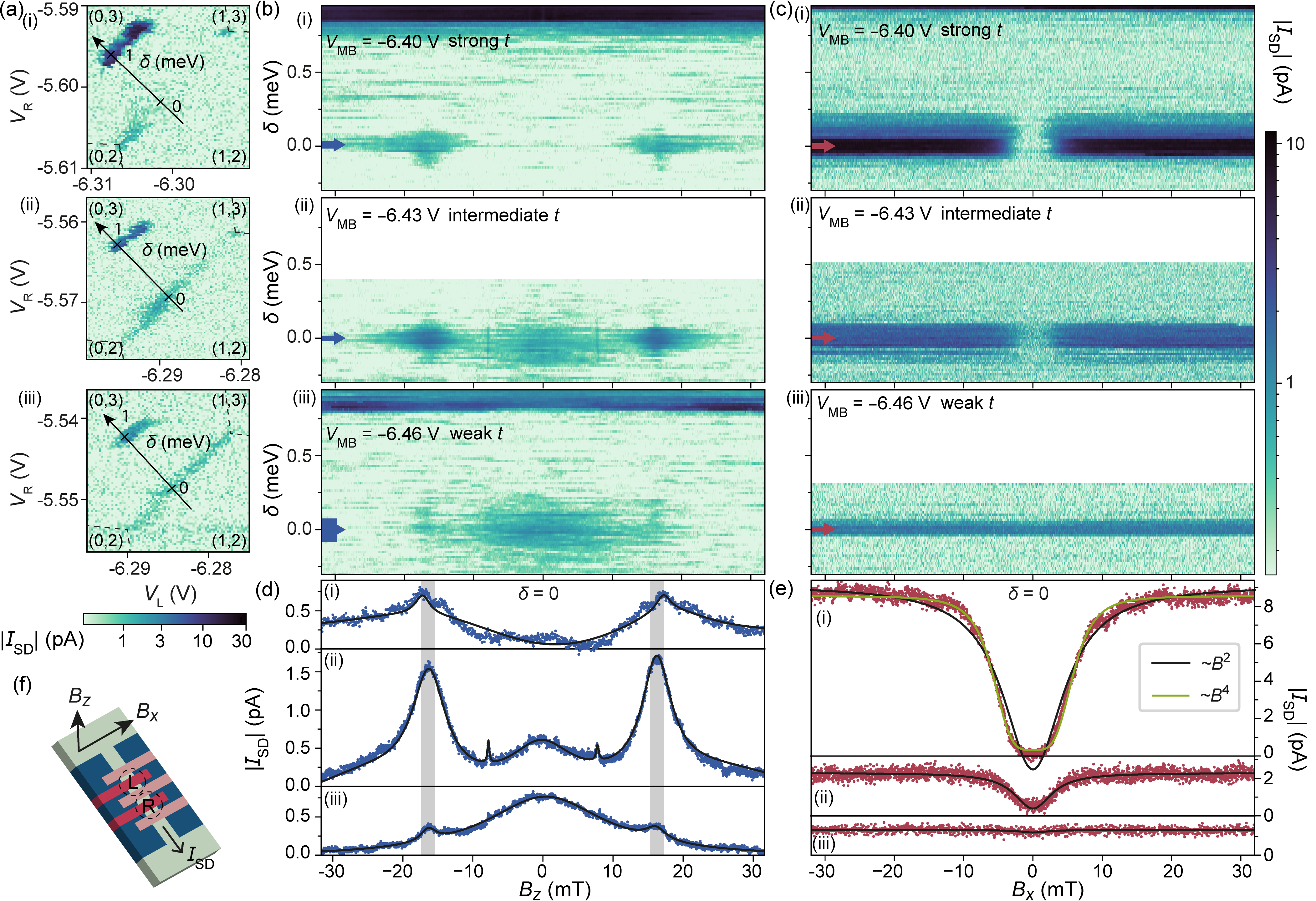}
	\caption{(a) Bias triangles with $V_\mathrm{SD}=\SI{-1}{mV}$ for the spin-blockaded transition $(1,2)\rightarrow(0,3)$ at zero magnetic field at (i) strong, (ii) intermediate, and (iii) weak interdot coupling $t$, at $V_\mathrm{MB}=\SI{-6.40}{V}$, $\SI{-6.43}{V}$, and $\SI{-6.46}{V}$, respectively, with corresponding maps as functions of $\delta$ [labeled in (a)] and $B$-field for (b) out-of-plane field $B_z$, and (c) in-plane field $B_x$. The field orientations are indicated in (f). (d,e): Line-cuts from the respective maps averaged around zero detuning over a range [indicated by arrows in (b,c)] of $\Delta\delta\approx\SI{30}{\micro eV}$ [$\approx\SI{150}{\micro eV}$ for (d,iii) due to weakness and instability of the signal] with fitted curves. A $B$-field offset is subtracted from each map and trace, assuming for (b) and (d): the side peaks are symmetric in $B_z$, and for (c) and (e): the zero-field dip is centered at $B_x = \SI{0}{T}$. Traces in (d) are fitted with multiple Lorentzians. In (e), the black traces show fits of Lorentzian-shaped dips, whereas the green trace in (i) is fitted empirically with higher order field dependence $B^4$.}
	\label{figlist}
\end{figure*}

The level structure of the $(1,2)$ states resulting from \eqref{eq:ham} is sketched in Fig.~\ref{figlevels}. At zero field, the Kramers pairs on the left dot are split by $\Delta_\mathrm{SO}$. A finite in-plane field $B_{x}$ (left side of the plot) does not couple to the valley degree of freedom, but aligns the spin-triplets in the right dot along the $x$-axis and splits them by the Zeeman energy. In an out-of-plane field $B_z$ (right side of the plot), both spin and valley states are split by $g\mu_B B_z$ with their respective $g_s$ and $g_v$, forming four clusters of states: one cluster of triplet states in the right dot per state in the left dot. The four fully blocked $S_z=\pm3/2$ states are sketched in red; all other states are open and sketched in yellow.

We now examine experimentally the dependence of spin blockade leakage current on the magnitude and orientation of an external magnetic field, and on the strength of interdot tunnel coupling $t$. The field direction is changed by rotating the sample, and $t$ is tuned via the middle barrier gate voltage, with weaker $t$ at more negative $V_\mathrm{MB}$. In Fig.~\ref{figlist}(a,i--iii) we show resulting bias triangles at zero field for various $t$, all weaker than in Fig.~\ref{figstart}(a). The corresponding current measured along the detuning axis $\delta$ and its dependence on out-of-plane $B_z$ and in-plane $B_x$ magnetic field is mapped in Fig.~\ref{figlist}(b,c). Line-cuts around $\delta=0$ averaged over a range of $\Delta\delta\approx\SI{30}{\micro eV}$ [$\approx\SI{150}{\micro eV}$ for Fig.~\ref{figlist}(d,iii) due to weakness and instability of the signal] are plotted as a function of $B_z$ and $B_x$ in Fig.~\ref{figlist}(d) and (e). 

Upon suppressing $t$, we observe in the bias triangles [Fig.~\ref{figlist}(a)] the emergence of a resonance around $\delta=0$ with current increasing from (i) below the noise level, to (iii) $\sim\SI{1}{pA}$. When turning on $B_z$ [Fig.~\ref{figlist}(b,d)], the current of this resonance decreases with increasing $B_z$ in a Lorentzian shape [Fig.~\ref{figlist}(b and d,ii and iii)], with full-width-half-maximum (FWHM) $\Delta B_z \sim \SI{10}{mT}$.

This zero-field current peak could arise from mixing of the blocked and the unblocked states by spin- or valley-mixing processes. As shown in Fig.~\ref{figlevels}, around zero field the four blocked states can undergo either (i) simultaneous onsite spin and valley flip in the one-carrier left dot, mixing e.g., the blocked state $\ket{\uparrow K^-}_L\ket{T^+_sS_v}_R$ with the unblocked $\ket{\downarrow K^+}_L\ket{T^+_sS_v}_R$, or (ii) onsite pure spin flips in the two-carrier right dot, mixing e.g., the blocked state $\ket{\uparrow K^-}_L\ket{T^+_sS_v}_R$ with the unblocked $\ket{\uparrow K^-}_L\ket{T^0_sS_v}_R$. The energy scale of the mixing terms competes with the external magnetic field: At large enough fields such that the energy splitting between the mixed states [$g_v\mu_{\mathrm B}B_z$ for (i) and $g_s\mu_{\rm B}B_z$ for (ii)] is larger than the mixing energy, the system will go into full spin blockade. The width of the zero-field current peak, therefore, contains information about the strength of the underlying mixing processes.

In double-dot systems for which similar zero-field peaks have been observed, the spin-mixing term has been attributed to the hyperfine interaction with randomly fluctuating nuclear spin baths~\cite{jouravlev2006electron, koppens2005control, CNTHFNatPhys, mittag2021few,yazyev2008hyperfine, fischer2009hyperfine}. If the same applies to our system and lifts the blockade via pure spin-flip processes (ii), then the peak width $\Delta B \sim \SI{10}{mT}$ should correspond to the root-mean-square magnitude of the random nuclear fields $B_\mathrm{nuc}$ experienced by the localized spins. Our quantum dots $\sim\SI{50}{nm}$ in radius are made of exfoliated BLG which contains only $1.01\%$ spinful $^{13}$C, yielding $B_{\rm nuc} \sim \SI{1}{\mu T}$ using the hyperfine coupling constant $A\sim 1$~$\mu$eV calculated by Refs.~\onlinecite{yazyev2008hyperfine, fischer2009hyperfine} for graphene, or $B_{\rm nuc}\sim \SI{100}{\mu T}$ using $A\sim 100$--$200~\mu$eV extracted from the leakage current peak observed in $^{13}$C enriched carbon nanotubes~\cite{CNTHFNatPhys}. Our observed peak width of $\SI{10}{mT}$ is, however, orders of magnitude larger, indicating that a different mechanism is responsible for this peak.

Wide zero-field peaks in Si have been attributed to finite-temperature co-tunneling effects, yielding $I\propto g_s\mu_\mathrm{B}B_z/\sinh(g_s\mu_\mathrm{B}B_z/k_\mathrm{B}T)$ when $t < k_{\rm B}T$~\cite{PhysRevB.86.115322,lai2011pauli,PhysRevLett.102.176806}. For type-(ii) processes where spins are flipped in the right dot, co-tunneling events have to involve virtual $(1,1)$ states that are too high in energy to be accessible. For type-(i) processes, virtual $(0,2)$ states are closer in energy and co-tunneling events with the lead can provide simultaneous spin- and valley-flips in the left dot; with $g_v=30$ and $T\approx\SI{100}{mK}$ we estimate for co-tunneling-induced current peaks FWHM~$\SI{12}{mT}$, similar to the measured $\SI{10}{mT}$. The skewness of the baseline resonance of the bias triangles towards the $(0,2)$ charge state in Fig.~\ref{figlist}(a,ii) corroborates with the conjecture, though noise and charge instability preclude definite conclusions. Another possible candidate could be site-specific spin--orbital effects specific to the symmetry of BLG~\cite{Guinea_2010}, leading to misaligned quantization axes along which $\Delta_{\rm SO}$ acts in the two dots, mixing spin and valley.

Another feature observed in out-of-plane field is two side peaks with FWHM~$\approx\SI{2}{mT}$ occurring at $B_z \approx \pm\SI{16.5}{mT}$, too narrow to arise from the co-tunneling effects discussed above. Considering the level structure in $B_z$ (see Fig.~\ref{figlevels}), we see that a cluster of triplet states with $\ket{\uparrow K^+}_L$ (moving up in energy) crosses the triplet with $\ket{\uparrow K^-}_L$ (moving down) at $g_v \mu_{\rm B} B_z \approx \Delta_{\rm SO}$ (gray shading in Fig.~\ref{figlevels}). If finite mixing between a blocked state in one triplet and an unblocked state in the other triplet exists, by a mechanism that flips valley in the left and spin in the right dot, e.g., between the blocked $\ket{\uparrow K^+}_L\ket{T^+_sS_v}_R$ and the unblocked $\ket{\uparrow K^-}_L\ket{T^0_sS_v}_R$, then an increase in current close to the crossing point would be expected. With $g_v \approx 30$ and side peaks at $\pm\SI{16.5}{mT}$, this implies $\Delta_{\rm SO} \approx \SI{30}{\micro eV}$, similar to that reported in Refs.~\onlinecite{AnnikaKondo, Luca1001} within a factor of two. 


We now turn to the effect of an in-plane magnetic field $B_x$, which we apply perpendicular to the double dot axis [see Fig.~\ref{figlist}(f)]. At strong interdot coupling $t$ [Fig.~\ref{figlist}(c,i) and (e,i)], we observe a large leakage current at finite $B_x$, which is reduced at zero field. As $t$ is weakened, the saturation leakage current decreases from $\SI{8}{pA}$ in (i) to $\SI{1.2}{pA}$ in (iii), while the width of the dip remains roughly constant.

In two-carrier spin blockade, observation of such a zero-field dip is usually an indication of strong spin--orbit interaction~\cite{danon2009pauli,pfund2007suppression,nadj-perge2010,Li2015,PhysRevB.86.115322,CNTHFNatPhys}. A similar mechanism can be expected to be at work here: At zero magnetic field, the level structure is dominated by the Kane--Mele spin--orbit interaction and has all quantization axes oriented along the $z$-axis, splitting the $(1,2)$ states into two clusters of six degenerate states (see Fig.~\ref{figlevels})---the system is in spin blockade. A finite $B_x$ lifts this blockade: not only does it lift the degeneracy, but it also tilts the quantization axis of the spin triplets towards the $x$-axis, such that the original blocked $S_z=\pm3/2$ states become mixed by the Zeeman field with the open states (the dark blocked states brightening in Fig.~\ref{figlevels}). As a consequence, a dip in the leakage current occurs around $B_x=0$. The leakage current weakens but remains finite when tilting the external magnetic field away from in-plane (see Appendix.~A3).

Details of the observed dip deviate from this simple interpretation. First of all, a direct mixing of blocked and unblocked states by $B_x$ should yield a regular Lorentzian line shape with a field dependence $B_x^2$~\cite{pfund2007suppression, nadj-perge2010, Li2015, PhysRevB.86.115322, CNTHFNatPhys}, which does not fit our data very well [see Fig.~\ref{figlist}(e,i), black]. In fact, an empirical fit with $A - C/(B_0^4 + B_x^4)$ where $A$ and $C$ are fitting constants, shows much better agreement [Fig.~\ref{figlist}(e,i), green], suggesting that the state-mixing by the applied Zeeman field is a higher-order effect. Secondly, the dip-like line shape is expected to arise from the competition between the Zeeman splitting and the interdot exchange energy $\propto t^2$ caused by coupling to the $(0,3)$ states; thus we expect the width of the dip $\propto t^2$~\cite{danon2009pauli}. The line shapes here, however, seem not to be strongly dependent on $t$. These two oddities suggest the existence of other mechanisms at play. The co-tunneling process discussed earlier in out-of-plane field cannot be responsible for the odd line shape, as here it is only relevant at much higher in-plane field since $g_s \ll g_v$.


To conclude, we examined the spin-blockade leakage current in a BLG double quantum dot at the three-carrier charge transition $(1,2)\rightarrow(0,3)$, and investigated its dependence on interdot coupling, and magnitude and orientation of external magnetic field. Most of the characteristics can be understood in terms of processes similar to those observed in other quantum dot systems, but some of the details of the underlying mechanisms still require further investigation. In out-of-plane magnetic field, the dominant feature observed is a zero-field current peak that could arise from finite-temperature co-tunneling with the leads, though explanation of the persistent sharp side peaks have to involve other mixing mechanisms. In in-plane magnetic field, we observe a zero-field current dip, which is expected based on the competition between the Zeeman effect and the Kane--Mele spin--orbit interaction; several details of its line shape, however, suggest additional mechanisms are at play. We expect further studies to capture the nature and strength of the various elusive spin-mixing mechanisms existing in BLG in more detail, thereby not only paving the way for BLG spin qubits, but also gaining deeper insights into BLG spin and valley physics.

\section*{acknowledgments}

We thank P. Märki and T. Bähler as well as the FIRST staff for their technical support. We acknowledge funding from the Core3 European Graphene Flagship Project, the Swiss National Science Foundation via NCCR Quantum Science and Technology, and the EU Spin-Nano RTN network. R. Garreis acknowledges funding from the European Union’s Horizon 2020 research and innovation programme under the Marie Skłodowska-Curie Grant Agreement No. 766025. K. W. and T. T. acknowledge support from the Elemental Strategy Initiative conducted by the MEXT, Japan, Grant Number JPMXP0112101001, JSPS KAKENHI Grant Number 19H05790 and JP20H00354.

\section*{Data availability}
The data supporting the findings of this study is made available via the ETH Research Collection.

\newpage

\setcounter{section}{0} 

\renewcommand\thesection{S~\arabic{section}} 
\setcounter{figure}{0} 
\renewcommand\thefigure{S\arabic{figure}}

\setcounter{section}{0} 
\renewcommand\thesection{S~\arabic{section}} 

\setcounter{figure}{0}
\renewcommand\thefigure{S\arabic{figure}}

\section{A1.~Methods}

\label{section:methods}

\begin{figure}
	\includegraphics[width=8.5cm]{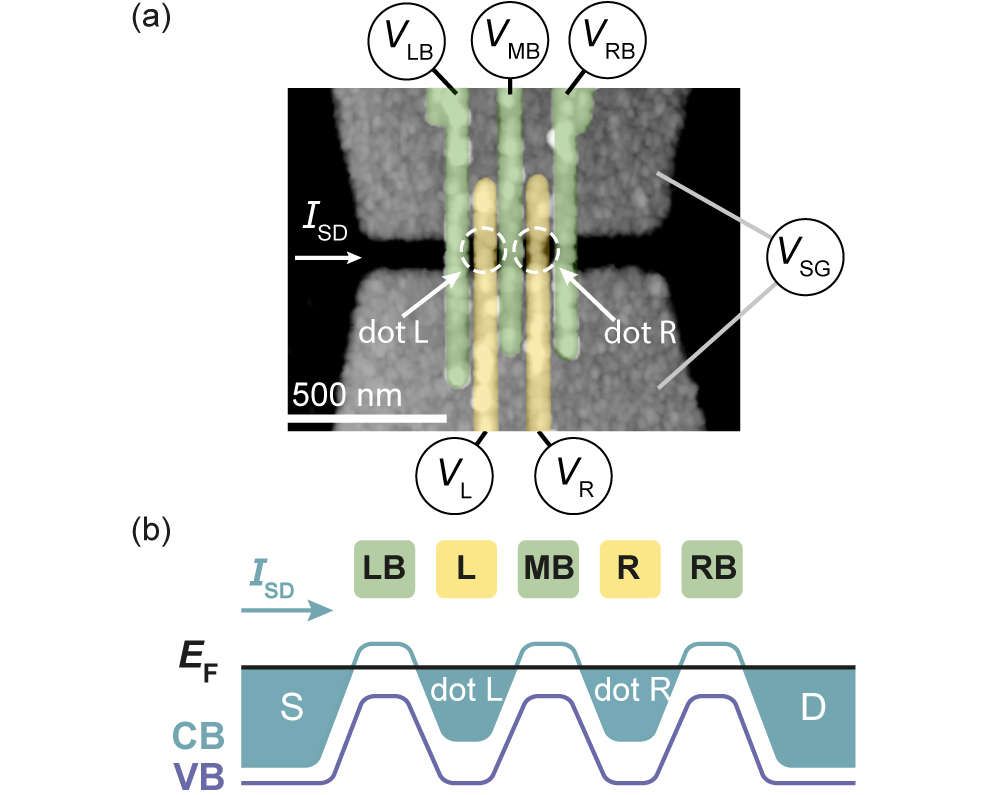}
	\caption{Reproduced from Supplementary information of Ref.~\cite{Blockade}: (a) False-colored AFM image of the sample used, where voltages supplied to these gates are labeled. (b) Schematic sketch of the conduction band (CB) and valence band (VB) edge variation along the channel, underneath the respective gates. The left and right dot, L and R, are formed underneath the respective plunger gates (yellow), with gate voltages $V_\mathrm{L}$, and $V_\mathrm{R}$. The two dot-lead tunnel couplings are controlled by left and right barrier gate voltages $V_\mathrm{LB}$ and $V_\mathrm{RB}$. The interdot tunnel coupling is controlled by the middle barrier gate voltage $V_\mathrm{MB}$.}
	\label{sampleS}
\end{figure}

The device is fabricated as described in Ref.~\onlinecite{Hiske2018electrostatically, MariusPRX, Blockade,Tunabledot}. A false-colored AFM image of the sample is shown in Fig.~\ref{sampleS}(a). Stacked with the dry-transfer technique~\cite{wang2013drytransferedge}, the van der Waals hetero-structure lies on a silicon chip with $\SI{280}{nm}$ surface SiO$_2$. The stack consists of a bottom graphite back gate, and on top of it a BLG flake encapsulated in $\SI{38}{nm}$ thick bottom and $\SI{20}{nm}$ thick top hBN flakes. Ohmic edge contacts with Cr and Au of $\SI{10}{}$ and $\SI{60}{nm}$ thickness, respectively, are evaporated after etching through the top hBN flake with reactive ion etching. A pair of $\SI{5}{nm}$ thick Cr, $\SI{20}{nm}$ thick Au split gates are deposited on top, defining a $\SI{1}{\micro m}$ long, $\SI{100}{nm}$ wide channel. Separated by a layer of $\SI{30}{nm}$ thick amorphous Al$_2$O$_3$ grown by atomic layer deposition, finger gates (yellow and green in Fig.~\ref{sampleS}) of $\SI{20}{nm}$ in width, and $\SI{5}{nm}$ Cr and $\SI{20}{nm}$ Au in thickness, lie across the channel. Neighboring finger gates are separated by $\SI{75}{nm}$ from center to center.

Our double quantum dots are defined electrostatically in the same BLG device as described in Refs.~\onlinecite{Blockade}. In BLG, a band-gap is formed near the $K^\pm$ valleys when applying a displacement field perpendicular to the sheet of BLG \cite{Ohta2006BLGband,mccann2006BLGband,Oostinga2008BLG}, where the size of the gap increases with the strength of the displacement field. With dual-gating, we therefore have control to both the size of the gap, and the doping, in the gated region. 

To form the pair of double electron dots studied in this work, we apply a positive global graphite back-gate voltage $V_\mathrm{BG}=\SI{5}{V}$, tuning the whole sheet of BLG into an $n$-doped regime. With negative voltages $V_\mathrm{SG}=\SI{-3.645}{V}, \SI{-3.53}{V}$ applied to the split gates [gray in Fig.~\ref{sampleS}(a)], we open up a band gap underneath the split gates, and simultaneously tune the Fermi energy $E_\mathrm{F}$ into the middle of the gap. Thus we form an $n$-type 1D-channel as shown in Fig.~\ref{sampleS}. With another layer of finger gates deposited on top, we gain control locally over the potential landscape within this channel. As shown in Fig.~\ref{sampleS}(b), we use the three barrier gates (green) LB, MB, and RB to tune the region underneath them into the gap, forming our left, middle, and right barriers. The barrier gate voltages control the location of the Fermi energy $E_\mathrm{F}$ in the gap, and hence the strength of the barriers. In general, the tunnel coupling decreases with more negative barrier gate voltages, until the voltages are negative enough for the formation of $p$-type dots underneath the barriers. The gates L and R serve as plunger gates for the electron dot L and R, respectively. With more negative plunger gate voltages $V_\mathrm{L}$ and $V_\mathrm{R}$, we can deplete the electron dots cleanly to the last carrier. 
\begin{figure}
	\includegraphics[width=8.5cm]{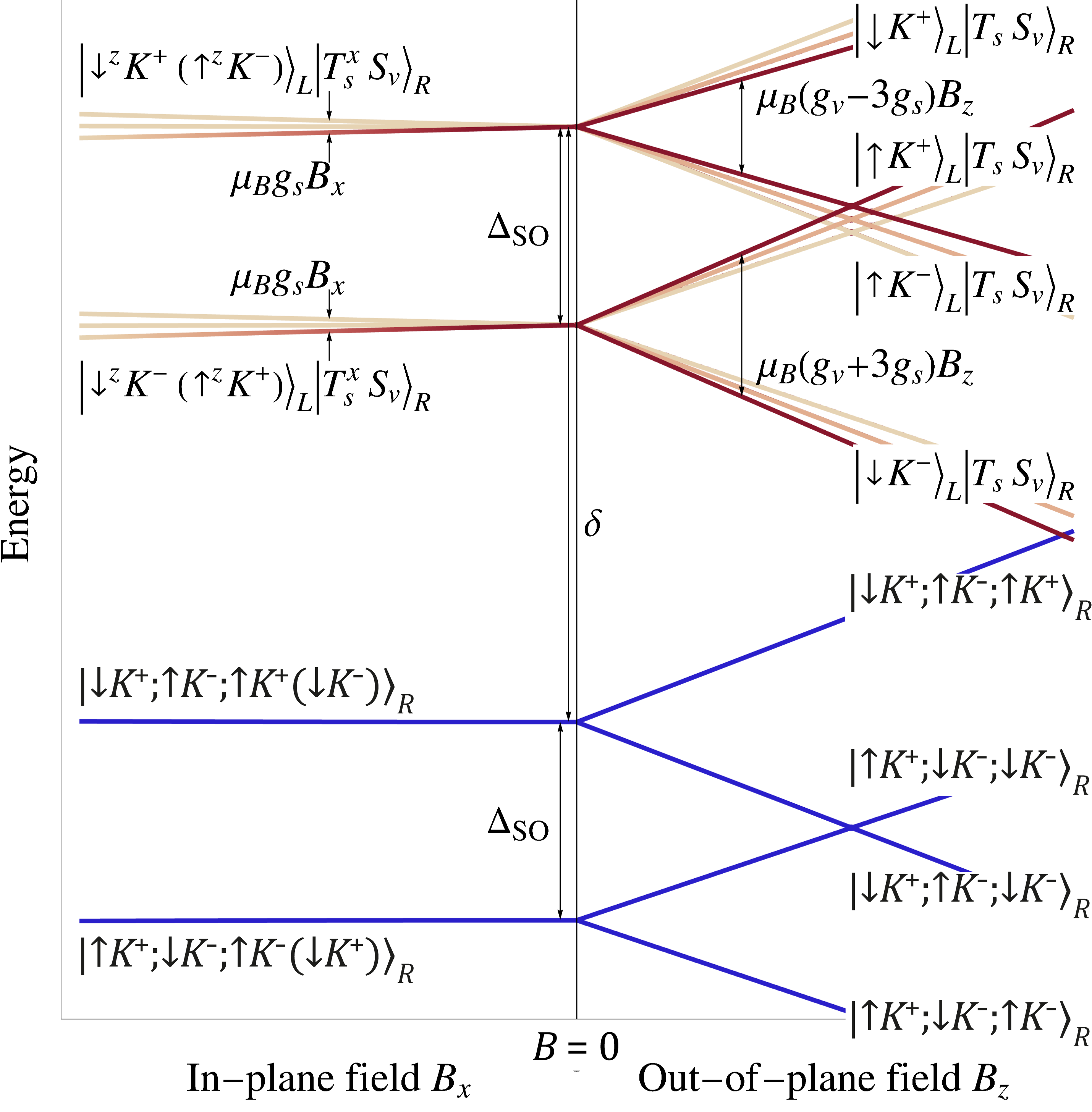}
	\caption{Sketch of the level diagram as a function of the magnetic field, assuming $\delta\gg t$ for clarity. The $(0,3)$ states are dark blue, in the $(1,2)$ subspace the color scale indicates each state's contribution to the transport, the lighter the color the larger the coupling to $(0,3)$. 
    For $B=0$ the dark red states $|Q_{\pm 3/2},\pm\rangle$ are fully blocked, while the other states are unblocked. Spin- and valley-mixing interactions can lift the blockade around $B =0$ and in the ``crossing'' region where the valley Zeeman splitting due to $B_z$ equals $\Delta_\mathrm{SO}$.
    For increasing $B_x$ the blockade is lifted due to a tilting of the spin-triplet quantization axis on the right dot.}
	\label{figleveldetail}
\end{figure}

The measurements are performed in a dilution refrigerator with base temperature of $\sim\SI{100}{\milli K}$. 

\section{A2.~Relevant (1,2) and (0,3) states}

Following the discussion in the main text, we specify here in more detail the states involved in transport.
The allowed $(1,2)$ charge states have two carriers in the right dot in a spin-triplet valley-singlet state and a single carrier in the left dot in any of the four spin-doublet valley-doublet states.
Thus, there are twelve possible $(1,2)$ states in total, consisting of one spin-quadruplet valley-doublet (eight states, all with total spin $S=3/2$) 
\begin{align*}
     |Q_{\frac{3}{2}},\tau\rangle = {} & {}  |{\uparrow \tau}\rangle_L |T_s^+ S_v\rangle_R ,\\
     |Q_{\frac{1}{2}},\tau\rangle = {} & {}  \frac{1}{\sqrt 3}\left(\sqrt{2}|{\uparrow \tau}\rangle_L |T_s^0 S_v\rangle_R +|{\downarrow \tau}\rangle_L |T_s^+ S_v\rangle_R \right),\\
     |Q_{-\frac{1}{2}},\tau\rangle = {} & {}  \frac{1}{\sqrt 3}\left(\sqrt{2}|{\downarrow \tau}\rangle_L |T_s^0 S_v\rangle_R +|{\uparrow \tau}\rangle_L |T_s^- S_v\rangle_R \right) ,\\
     |Q_{-\frac{3}{2}},\tau\rangle = {} & {}  |{\downarrow \tau}\rangle_L |T_s^- S_v\rangle_R ,
\end{align*}
and one spin-doublet valley-doublet (four states with $S=1/2$)
\begin{align*}
    |D_{\frac{1}{2}},\tau\rangle = {} & {}  \frac{1}{\sqrt 3}\left(|{\uparrow \tau}\rangle_L |T_s^0 S_v\rangle_R -\sqrt{2}|{\downarrow \tau}\rangle_L |T_s^+ S_v\rangle_R \right),\\
     |D_{-\frac{1}{2}},\tau\rangle = {} & {}  \frac{1}{\sqrt 3}\left(|{\downarrow \tau}\rangle_L |T_s^0 S_v\rangle_R -\sqrt{2}|{\uparrow \tau}\rangle_L |T_s^- S_v\rangle_R \right),
 \end{align*}
where $\tau = \pm$ denotes the valley quantum number $K^\pm$, $T_s^{\pm,0}$ are the spin triplet states, $S_v$ the valley singlet and the index $L$ ($R$) indicates the left (right) QD~\cite{Buchachenko2002,PhysRevB.82.075403}.
As explained in the main text, the four $(0,3)$ states are all spin-doublet states, meaning that in this basis only four out of twelve states would be coupled to them and thus be unblocked.

The spin quadruplet and doublet states specified above are not eigenstates of the Hamiltonian given in Eq.~(1) of the main text.
In the presence of an out-of-plane field $B_z$ and Kane--Mele spin--orbit coupling one finds the eigenstates $|{s\tau}\rangle_L |T_s S_v\rangle$, as used in the main text, with $s = \uparrow,\downarrow$ labeling spin along $\hat z$.
Written in terms of the basis states given above, these read as
\begin{align}
    |{\uparrow \tau}\rangle_L |T_s^+ S_v\rangle_R {} & {} =  |Q_{\frac{3}{2},\tau}\rangle,\label{eq:s1}\\
    |{\uparrow \tau}\rangle_L |T_s^0 S_v\rangle_R {} & {} =  \frac{1}{\sqrt 3} \left( \sqrt 2 |Q_{\frac{1}{2},\tau}\rangle + |D_{\frac{1}{2},\tau}\rangle \right) ,\\
    |{\uparrow \tau}\rangle_L |T_s^- S_v\rangle_R {} & {} =  \frac{1}{\sqrt 3} \left( |Q_{-\frac{1}{2},\tau}\rangle -\sqrt 2 |D_{-\frac{1}{2},\tau}\rangle \right) ,\\
    |{\downarrow \tau}\rangle_L |T_s^+ S_v\rangle_R {} & {} = \frac{1}{\sqrt 3} \left( |Q_{\frac{1}{2},\tau}\rangle -\sqrt 2 |D_{\frac{1}{2},\tau}\rangle \right),\\
    |{\downarrow \tau}\rangle_L |T_s^0 S_v\rangle_R {} & {} = \frac{1}{\sqrt 3} \left( \sqrt 2 |Q_{-\frac{1}{2},\tau}\rangle + |D_{-\frac{1}{2},\tau}\rangle \right),\\
    |{\downarrow \tau}\rangle_L |T_s^- S_v\rangle_R {} & {} = |Q_{-\frac{3}{2}},\tau\rangle.\label{eq:s6}
\end{align}
We thus see that in the $S_z = \pm 1/2$ subspace the blocked and unblocked states are mixed, leaving only the four states $|Q_{\pm\frac{3}{2}},\pm\rangle$ as truly blocked states.
If the magnetic applied field is \emph{in-plane} instead of out-of-plane, the Zeeman effect will mostly result in a rotation of the three spin-triplet states on the right dot, aligning their spin quantization axis along $\hat x$ (as long as $g_s\mu_{\rm B}B_x \ll \Delta_{\rm SO}$ the Kramers pair on the left dot will not be affected much).
This results in a mixing of all three states $T_s^{\pm,0}$ in each spin-triplet in (\ref{eq:s1}--\ref{eq:s6}) (which were defined along $\hat z$), lifting the blockade for all twelve states.

Including the $(0,3)$ states and adding a finite detuning $\delta$ between the $(1,2)$ and $(0,3)$ states for clarity, we show a typical resulting level diagram in Fig.~\ref{figleveldetail}. The $(0,3)$ states are dark blue, in the $(1,2)$ subspace the color scale indicates each state's contribution to the transport, the lighter the color the larger the coupling to $(0,3)$. At zero field the Kramers pairs are split by $\Delta_\mathrm{SO}$. With an out-of-plane field $B_z$ the spins are split by the Zeeman splitting $g_s \mu_B B_z$ and the valleys are split by $g_v \mu_B B_z$. The states with $S_z = \pm 3/2$ are blocked, while the other states are open to the transport.

One subtlety we would like to point out here is that without spin- or valley-flip tunneling the $(1,2)$ states that have a $T_s^0$ spin configuration on the right dot (i.e., states $|s \tau\rangle_L |T_s^0 S_v\rangle_R$, with $s=\uparrow,\downarrow$ and $\tau=\pm$), couple only to $(0,3)$ states in a Kramers pair with opposite parity, i.e., split off by $\Delta_\mathrm{SO}$, which makes their contribution to the transport inefficient at $\delta=0$.
However, if the decay rate $\Gamma$ of the $(0,3)$ states to the right lead is sufficiently large, $\Gamma \gg \Delta_\mathrm{SO}$, all unblocked states can be treated as equally open, nonetheless.

\section{A3.~Dependence of zero-field dip on magnetic field orientation}

\label{section:30deg}

\begin{figure}
	\includegraphics[width=8.5cm]{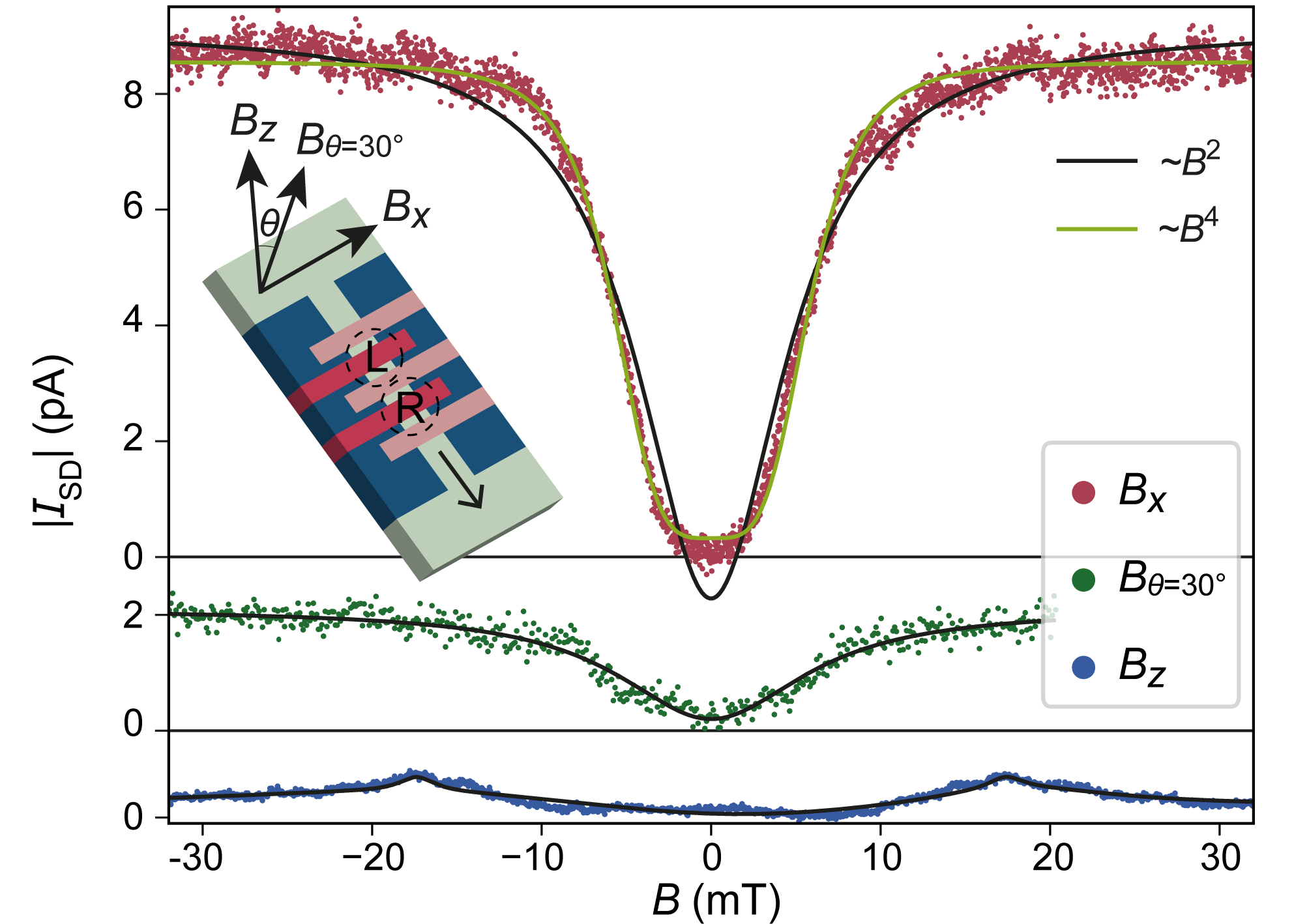}
	\caption{Current $|I_\mathrm{SD}|$ measured around $\delta=0$ as a function of external magnetic field $B$ applied out-of-plane, at $30^{\circ}$ with the out-of-plane direction, and in-plane (across the channel), at strong interdot coupling $t$ with $V_\mathrm{MB}=\SI{-6.40}{V}$. The blue and the red traces are the same as the traces in Fig.~3(d,i) and (e,i) in the main text. Inset: schematics of the sample with the $B$-field directions labeled. Black line: fit with a Lorentzian dip; green line: fit with a ``Lorentzian'' with a $\sim B^4$-dependence.}
	\label{fig30deg}
\end{figure}

The leakage current dips at strong interdot coupling at three different magnetic field orientations are plotted together in Fig.~\ref{fig30deg}. The blue and red traces, for out-of-plane magnetic field $B_z$ and for in-plane magnetic field $B_x$, are the same as the data shown in Fig.~3(d,i) and (e,i) in the main text. When we tilt the external field away from the in-plane direction, we see that at $B_{\theta=30^{\circ}}$ (shown in inset in Fig.~\ref{fig30deg}, the saturation current at large field reduces from $\SI{8}{pA}$ in in-plane field, to $\SI{2}{pA}$. Finally, when the field direction is completely out-of-plane in $B_z$, this zero field dip feature seems to disappear completely.

This magnetic field orientation dependence corroborates with the assumption in the main text that a zero-field spin--orbit interaction aligned along the $z$-axis, i.e., that the Kane--Mele type spin--orbit gap aligning and spins and valleys along the $z$-axis, is at work here.

\section{A4.~valley blockade leakage current}

\label{section:VB1120}
\begin{figure}
	\includegraphics[width=8.5cm]{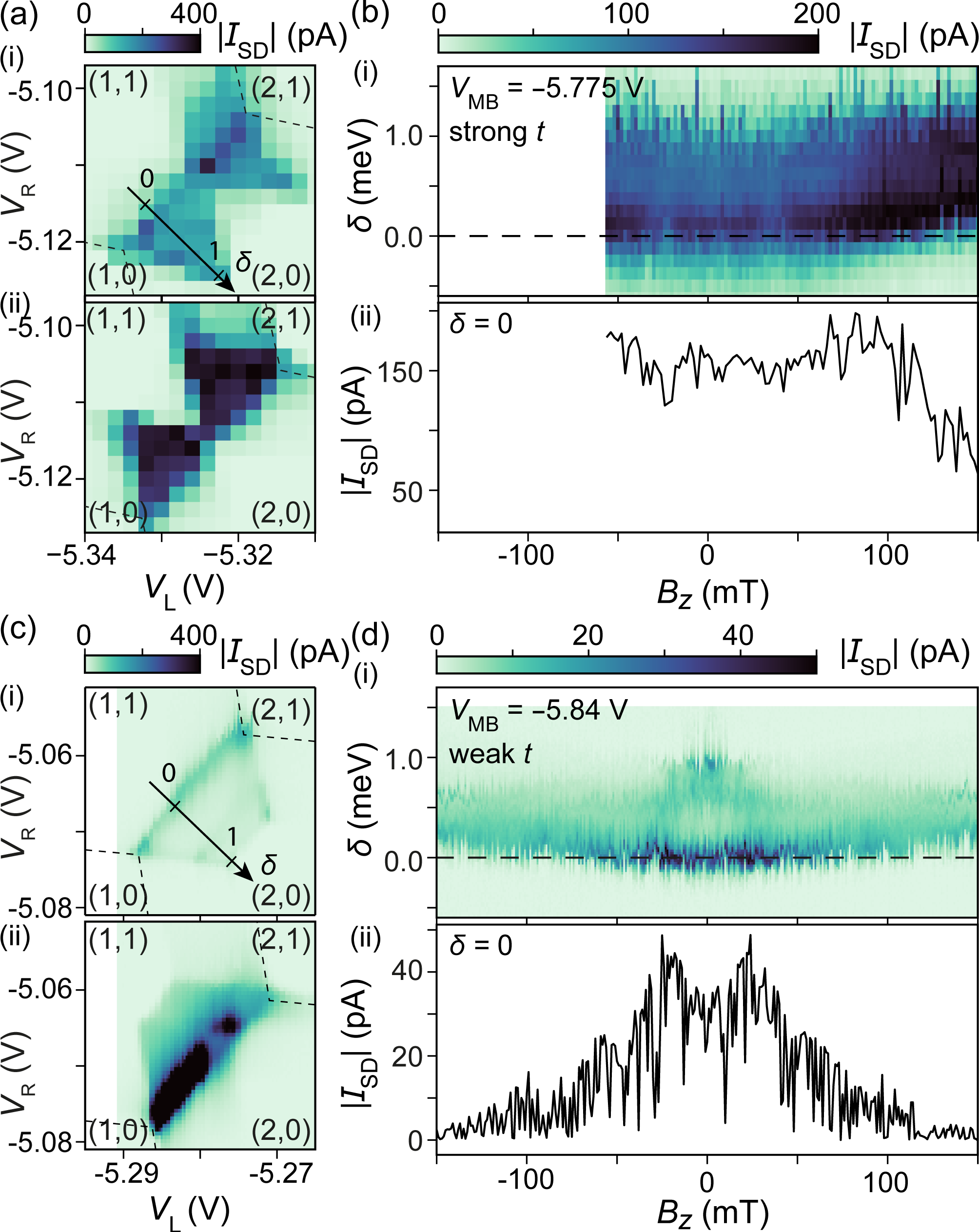}
	\caption{Finite-bias triangles near the two-electron transition $(1,1)\leftrightarrow(2,0)$, at (a) strong interdot coupling with $V_\mathrm{MB}=\SI{-5.775}{V}$, and (c) weak interdot coupling with $V_\mathrm{MB}=\SI{-5.84}{V}$, at (i) positive source-drain bias $V_\mathrm{SD}=\SI{1}{mV}$, and (ii) negative source-drain bias $V_\mathrm{SD}=\SI{-1}{mV}$. For negative bias, the transition $(1,1)\rightarrow(2,0)$ is Pauli valley blocked.
	(b,d) The perpendicular magnetic field dependence of a line-cut around the detuning-axis $\delta$, marked in (a,c). The detuning is converted from the left and the right plunger gate voltages $V_\mathrm{L}$ and $V_\mathrm{R}$, where in (i) the entire map and in (ii) a cut along $\delta=0$ is shown. A zero-field minimum is observed in (b) whereas in (d) a zero-field maximum is found.}
	\label{VB1120}
\end{figure}
Zero-field valley blockade in BLG double dots near two-carrier transition $(1,1)\leftrightarrow(2,0)$ has been observed and reported in Ref.~\cite{Blockade}. We examine the valley blockade leakage current at strong Fig.~\ref{VB1120}(a,b) and weak Fig.~\ref{VB1120}(c,d) interdot couplings, and see that the valley blockade leakage current also exhibits a dependence on the perpendicular magnetic field $B_z$. The bias triangles are shown in Fig.~\ref{VB1120}(a,c) for strong, and weak interdot couplings, where the bias is applied in direction of (i) the valley blockaded $(1,1)\rightarrow(2,0)$ transition, and (ii) the non-blockaded $(0,2)\rightarrow(1,1)$ transition. The magnetic field $B_z$ is swept along the detuning $\delta$-axis labelled on the triangles in (a) (c,i). At strong interdot coupling [Fig.~\ref{VB1120}(b)], the leakage current increases with magnetic field; at weak interdot coupling on the other hand [Fig.~\ref{VB1120}(d)], a zero-field peak is observed. The exact width of the current peaks and dips are hard to extract due to the bending of the $\delta=0$ base-line, more prominent than that for the $(1,2)\leftrightarrow(0,3)$ configuration, due to the valley $g$-factor $g_\mathrm{v}\approx 30$ being much larger than the spin $g$-factor $g_\mathrm{s}=2$ \cite{Tunabledot}.
\end{document}